\documentclass[a4paper]{article}
\usepackage{a4wide}
\usepackage[normalem]{ulem}
\usepackage{dsfont}
\usepackage[T1]{fontenc}
\usepackage{amsmath}
\usepackage{amssymb}
\usepackage{amsthm}
\usepackage[ansinew]{inputenc}
\usepackage{enumerate}
\usepackage{fancyhdr}
\usepackage{color}
\usepackage{ifpdf}
\usepackage{graphicx}
\usepackage{rotating}
\usepackage[T2A]{fontenc}
\usepackage{mathtext}
\usepackage{caption}
\DeclareMathOperator{\tr}{tr}

\theoremstyle{plain}

\newtheorem*{thm*}{Theorem}

\newtheorem*{prop*}{Proposition}
\newtheorem{cor}{Corollary}

\usepackage{authblk}

\renewcommand\Authands{ and }

\begin{document}
 \title{Isotropization of solutions of the Einstein-Vlasov system with Bianchi V symmetry}
\author[1,2,3]{Ernesto Nungesser\footnote{ernesto@maths.tcd.ie}}
\author[3]{Lars Andersson\footnote{laan@aei.mpg.de}}    
\author[4]{Soumyajit Bose\footnote{soumyab@iitk.ac.in}}
\author[5]{Alan A. Coley\footnote{aac@mathstat.dal.ca}}
\affil[1]{School of Mathematics, Trinity College, Dublin 2, Ireland}
\affil[2]{Department of Mathematics, Royal Institute of Technology, 10044~Stockholm, Sweden}
\affil[3]{Max-Planck-Institute for Gravitational Physics, Am~M\"{u}hlenberg~1, 14476~Golm, Germany}
\affil[4]{Indian Institute of Technology, Kanpur, Uttar Pradesh-208016, India}
\affil[5]{Department of Mathematics and Statistics, Dalhousie University, Halifax, Nova~Scotia, Canada B3H 3J5}

\renewcommand\Authands{ and }
\maketitle
\begin{abstract}
Using the methods developed for different Bianchi class A cosmological models we treat the simplest Bianchi class B model, namely Bianchi type V.
The future non-linear stability for solutions of the Einstein-Vlasov system is demonstrated and it is shown that these solutions are asymptotically stable to the Milne solution. 
Within the isotropic solutions of the Einstein-Vlasov system the spatially flat Friedmann solution is unstable within this class, and expanding models
tend also to the Milne solution.
\end{abstract}

 \section{Introduction}
 Recently strong results have been obtained concerning the non-linear stability of the spatially flat Friedmann solution coupled to collisionless matter
and a scalar field within the inhomogeneous class \cite{Hans}. These results rely on a positive cosmological constant or, at least,
on a scalar field which converges to a positive non-degenerate minimum of the potential. 

Without a cosmological constant such statements are much more difficult to obtain, since the associated
exponential behavior towards the future is absent. A natural strategy is to simplify the problem assuming symmetries. This was done for instance in \cite{EN} where non-linear stability was shown
without a cosmological constant, but assuming Bianchi I symmetry and assuming small data.  However, considering only isotropic solutions with a fluid without a cosmological constant it has been shown (for a proof cf. \cite{EW2}) that the flat Friedmann solution is unstable within the isotropic class
i.e., it tends to the solutions of $K=-1$ or $K=1$, where $K$ is the usual curvature parameter which we will define in Section \ref{sap}.  Here we establish the corresponding statement for collisionless matter.

The isotropic and forever expanding solutions tend to the Milne solution, which is an empty, isotropic universe with $K=-1$ and is actually simply a piece of Minkowski spacetime described in expanding coordinates (cf. P. 27 of \cite{Muk} for details). This solution is also the future attractor of solutions of the Einstein-Euler system with Bianchi V symmetry when considering a non-tilted fluid with a linear equation of state $P=(\gamma-1)\rho$ and $2/3 < \gamma < 2$, where $P$ is the pressure and $\rho$ the energy density of the fluid and $\gamma$ is a constant. It is thus natural to ask whether stability holds in the spirit of \cite{EN}. The physical idea is the following. Due to the expansion of the Universe it is reasonable to expect that the dispersion of the velocities of the particles (which in our case describe galaxies) tends to zero, such that the collisionless gas can be approximated by dust. We prove
that this is the case assuming small data. The exact meaning of small will be specified later, but basically
it means that one is initially arbitrarily close to the corresponding dust solution, which in particular implies that the initial dispersion of the velocities is small. 

When considering collisionless matter one cannot assume that the metric and the energy-momentum tensor are diagonal, due to the Vlasov equation. In order to obtain some intuition it is thus natural to also have a look at what happens in the case of tilted fluids. For the tilted fluid model the stability analysis, performed in the full state space for all of 
the equilibrium points, was presented in \cite{ColeyHervik,HVC}. The equilibrium points important for the late-time behavior in the Bianchi V invariant
space are the Milne solution, which is future stable  for $2/3 < \gamma < 4/3$, and the so-called extremely tilted Milne solution, which is future stable for $6/5 < \gamma  < 2$. Expanding spatially homogeneous tilted perfect fluid models were also studied in \cite{CHL,CHLM}.
It was shown that for ultra-radiative equations of state (i.e., $\gamma>4/3$), generically the tilt becomes extreme at late times and the fluid
observers will reach infinite expansion within a finite proper time and experience a singularity. 
In particular, for irrotational Bianchi type V perfect fluid models with $4/3 < \gamma < 2$, to the future the fluid world-lines become null (with respect
to the geometric congruence), and the expansion, the shear, the acceleration, and the length scale all diverge but the matter density (and curvature scalars) tend
to zero (i.e. the models end at a kinematic singularity) \cite{CHL,CHLM}.

One can see that the singular behavior appears for values of $\gamma$ far from dust (which corresponds to $\gamma=1$) and in fact we could show that in the case of consideration here no singular behavior occurs. We prove not only non-linear stability within the symmetry class, but asymptotic stability. In other words the solution not only stays close to the corresponding solution, but tends towards it. This means that assuming small data, solutions of the Einstein-Vlasov system with Bianchi V symmetry tend to the Milne solution and thus isotropize. 

Let us mention that the Milne solution is also a motivation for the weak global asymptotics conjecture of Anderson \cite{Anderson} which is related to the
geometrization of 3-manifolds. Moreover the Milne solution is stable under perturbations in the vacuum setting \cite{AM,AM2}. Ringstr\"{o}m has linked this conjecture
to the isotropization of the Universe as a consequence of the evolution, in the sense that the fraction of observers considering the universe to tend to become
isotropic tends to unity asymptotically \cite{Hans}. This idea contrasts with the result of Collins and Hawking who showed that for physically reasonable matter
the set of spatially homogeneous cosmological models which approach isotropy at infinite times is of measure zero in the space of all spatially homogeneous models \cite{CollinsH}.
We note that (generalized) $K=-1$ Friedmann-Lema{\^{\i}tre cosmologies are also of interest in the averaging problem \cite{Martin,CPZ}.
\section{The Einstein-Vlasov system}
A cosmological model represents a universe at a certain averaging scale. It is described via a Lorentzian metric $g_{\alpha\beta}$
 (we will use signature -- + + +) on a manifold $M$ and a family of fundamental observers. The metric is assumed to be time-orientable, which means that at each point of 
 $M$ the two halves of the light cone can be labeled past and future in a way which varies continuously from point to point. This enables us to distinguish between 
 future-pointing and past-pointing timelike vectors.
 The interaction between the geometry and the matter is described by the Einstein field equations (we use geometrized units i.e.
 the gravitational constant G and the speed of light in vacuum c are set equal to one):
\begin{eqnarray*}
G_{\alpha\beta}= 8\pi T_{\alpha \beta},
\end{eqnarray*}
where $G_{\alpha\beta}$ is the Einstein tensor and $T_{\alpha \beta}$ is the energy-momentum tensor. For the matter model
 we will take the point of view of kinetic theory \cite{St}. The sign conventions of \cite{RA} are used. Also the Einstein summation convention that repeated indices
 are to be summed over is used. Latin indices run from one to three and Greek ones from zero to three.

Consider a particle with non-zero rest mass which moves under the influence of the gravitational field. The mean field will be described now by the metric and the components 
of the metric connection. The worldline $x^\alpha$ of a particle  is a timelike curve in spacetime. The unit future-pointing tangent vector to this curve is the 4-velocity $v^{\alpha}$ and $p^{\alpha}=mv^{\alpha}$
 is the 4-momentum of the particle. Let $T_x$ be the tangent space at a point $x^{\alpha}$ in the spacetime $M$, then we define the
phase space for particles of arbitrary rest masses $P$ to be the following set:
\begin{eqnarray*}
P=\{(x^{\alpha},p^{\alpha}):\ x^{\alpha} \in M,\ p^{\alpha} \in T_xM,\ p_{\alpha} p^{\alpha} \leq 0,\ p^{0}>0\},
\end{eqnarray*}
which is a subset of the tangent bundle $TM=\{(x^{\alpha},p^{\alpha}):\ x^{\alpha} \in M,\ p^{\alpha} \in T_xM\}$.
For particles of the same type and with the same rest mass $m$ which is given by the mass shell relation:
\begin{eqnarray*}
 p_{\alpha} p^{\alpha}=-m^2,
\end{eqnarray*}
we have the phase space $P_m$ for particles of mass $m$:
\begin{eqnarray*}
P_m=\{(x^{\alpha},p^{\alpha}):\ x^{\alpha} \in M,\ p^{\alpha} \in T_xM,\ p_{\alpha} p^{\alpha}=-m^2,\ p^{0}>0\}.
\end{eqnarray*}

We will consider from now on that all the particles have unit mass. We have then that the possible values for the 4-momenta are all future pointing unit timelike vectors. 
These values form the hypersurface:
\begin{eqnarray*}
P_1=\{(x^{\alpha},p^{\alpha}):\ x^{\alpha} \in M,\ p^{\alpha} \in T_xM,\ p_{\alpha} p^{\alpha}= -1,\ p^{0}>0\},
\end{eqnarray*}
which we will call the mass shell. The collection of particles (galaxies or clusters of galaxies) will be described (statistically) by a non-negative real valued
 distribution function $f(x^\alpha,p^\alpha)$ on $P_1$. This function represents the density of particles at a given spacetime point with
 given four-momentum. A free particle travels along a geodesic. Consider now a future-directed timelike geodesic parametrized by proper time $s$.
 The tangent vector is then at any time future-pointing unit timelike. The equations of motion thus define a flow on $P_1$ which is generated by a vector field $L$ which is called geodesic
 spray or Liouville operator. The geodesic equations are:
\begin{eqnarray*}
\frac{dx^{\alpha}}{ds}=p^{\alpha}; \ \ \frac{dp^{\alpha}}{ds}=-\Gamma^\alpha_{\beta \gamma} p^{\beta} p^{\gamma},
\end{eqnarray*}
where the components of the metric connection, i.e. 
$\Gamma_{\alpha\beta \gamma}=g(e_{\alpha},\nabla_{\gamma}e_{\beta})=g_{\alpha\delta}\Gamma^{\delta}_{\beta\gamma}$ can be expressed in the
 vector basis $e_\alpha$ as [(1.10.3) of \cite{JS}]:
\begin{align}\label{con}
\Gamma_{\alpha\beta\gamma}=\frac12[e_{\beta}(g_{\alpha\gamma})+e_{\gamma}(g_{\beta\alpha})+e_{\alpha}(g_{\gamma\beta})+\eta^\delta_{\gamma\beta}g_{\alpha\delta}+\eta^{\delta}_{\alpha\gamma}g_{\beta\delta}-\eta^{\delta}_{\beta\alpha}g_{\gamma\delta}].
\end{align}
The commutator of the vectors $e_\alpha$ can be expressed with the following formula:
\begin{eqnarray*}
[e_{\alpha},e_{\beta}]=\eta^{\gamma}_{\alpha \beta}e_{\gamma},
\end{eqnarray*}
where $\eta^{\gamma}_{\alpha \beta}$ are called commutation functions.

The restriction of the Liouville operator to the mass shell using the geodesic equations has the following form:
\begin{align*}
 L=p^{\alpha}\frac{\partial}{\partial x^{\alpha}}-\Gamma^a_{\beta \gamma} p^{\beta} p^{\gamma}\frac{\partial}{\partial p^{a}},
\end{align*}
where now there is no dependence on $p^0$ anymore.
 We will consider the collisionless case which is described via the Vlasov equation:
\begin{align*}
L(f)=0.
\end{align*}
The unknowns of our system are a 4-manifold $M$, a Lorentz metric $g_{\alpha\beta}$ on this manifold and the distribution function $f$
 on the mass shell $P_1$ defined by the metric. We have the Vlasov equation defined by the metric for the distribution function and the
 Einstein equations. It remains to define the energy-momentum tensor $T_{\alpha\beta}$ in terms of the distribution and the metric. Before that we need a Lorentz 
 invariant volume element on the mass shell. A point of the tangent space has the volume element
$ |g^{(4)}|^{\frac{1}{2}} dp^0 dp^1 dp^2 dp^3$ ($g^{(4)}$ is the determinant of the spacetime metric) which is Lorentz invariant. Now considering $p^0$ as a dependent
variable the induced (Riemannian) volume of the mass shell considered as a hypersurface in the tangent space at that point is
\begin{eqnarray*}
 \varpi=2H(p^{\alpha})\delta( p_{\alpha} p^{\alpha}+1)|g^{(4)}|^{\frac{1}{2}} dp^0 dp^1 dp^2 dp^3,
\end{eqnarray*}
where $\delta$ is the Dirac distribution function and $H(p^{\alpha})$ is defined
 to be one if $p^{\alpha}$ is future directed and zero otherwise.
We can write this explicitly as:
 \begin{eqnarray*}
\varpi=|p_0|^{-1} |g^{(4)}|^{\frac{1}{2}} dp^1 dp^2 dp^3.
\end{eqnarray*}
Now we define the energy momentum tensor as follows:
 \begin{eqnarray*}
T_{\alpha\beta}=\int f(x^{\alpha},p^{a}) p_{\alpha}p_{\beta}\varpi.
\end{eqnarray*}
\textbf{Assumption}: We will assume that $f$ has compact support in momentum space for each fixed t. 

One can show that $T_{\alpha\beta}$ is divergence-free and thus it is compatible with the Einstein equations. For collisionless
 matter all the energy conditions hold (for details we refer to \cite{RIV}).  In particular the dominant energy condition is equivalent to the statement that
 in any orthonormal basis the energy density dominates the other components of $T_{\alpha\beta}$, i.e. $T_{\alpha\beta}\leq T_{00}$
 for each $\alpha,\beta$ (P. 91 of \cite{HaEl}). Using the mass shell relation one can see that this holds for collisionless matter.
 The non-negative sum pressures condition is in our case equivalent to $g_{ab}T^{ab} \ge 0$.
For forever expanding Bianchi models (Friedmann solutions are a subset) coupled to dust or to collisionless matter the spacetime is future complete (theorem 2.1 of \cite{GC}).
For a general introduction into the initial value problem we refer to \cite{RA,Hans}.

\section{Description of Bianchi spacetimes via the metric approach}
The only Bianchi spacetimes which admit a compact Cauchy hypersurface are Bianchi I and IX. 
In order to be not that restrictive we will consider locally spatially homogeneous spacetimes. They are defined as follows. 
Consider an initial data set on a three-dimensional manifold $S$. Then this initial data set is called locally spatially homogeneous 
if the naturally associated data set on the universal covering $\tilde{S}$ is homogeneous. For Bianchi models the universal covering space $\tilde{S}$
can be identified with its Lie group $G$. A Bianchi spacetime admits a Lie algebra of Killing vector fields with basis $\mathbf{k}_1$, $\mathbf{k}_2$, $\mathbf{k}_3$
and structure constants $C^{c}_{ab}$, such that:
\begin{eqnarray*}
[\mathbf{k}_a,\mathbf{k}_b]=-C^{c}_{ab}\mathbf{k}_c.
\end{eqnarray*}
The Killing vector fields $\mathbf{k}_a$ are tangent to the group orbits which are called surfaces of homogeneity. If one chooses a unit vector field $n$ normal
to the group orbits we have a natural choice for the time coordinate $t$ such that the group orbits are given by a constant $t$. 
This unit normal is invariant under the group. One can now choose a triad of spacelike vectors $e_{a}$ that are tangent to the group orbits 
and that commute with the Killing vector fields. A frame $\{\mathbf{n},\mathbf{e}_{a}\}$ chosen in this way is called a \textbf{left invariant frame} and
it is generated by the right invariant Killing vector fields. 
Since $\mathbf{n}$ is hypersurface orthogonal the vector fields
 $\mathbf{e}_a$ generate a Lie algebra with structure constants $\eta^c_{ab}$. It can be shown that this Lie algebra is in fact equivalent to the Lie algebra of the
 Killing vector fields. Thus one can classify the Bianchi spacetimes using either the structure constants or the spatial  commutation functions of the basis vectors. 
 The remaining freedom in the choice of the frame is a time-dependent linear transformation, which can be used to introduce a set of time-independent spatial vectors $\mathbf{E}_a$.
 The corresponding commutation functions are then constant in time and one can make them equal to the structure constants:
\begin{eqnarray*}
[\mathbf{E}_a,\mathbf{E}_b]=C^c_{ab}\mathbf{E}_c.
\end{eqnarray*}
If $\mathbf{W}^a$ denote the 1-forms dual to the frame vectors $\mathbf{E}_a$
the metric of a Bianchi spacetime takes the form:
\begin{eqnarray}\label{mt}
^4 g=-dt^2+g_{ab}(t)\mathbf{W}^a \mathbf{W}^b,
\end{eqnarray}
where $g_{ab}$ (and all other tensors) on $G$ will be described in terms of the frame components of the left invariant frame which has been introduced. The structure 
constants can be put in the following form, where $\nu$ is a symmetric matrix and the vector $(a_1,a_2,a_3)$ is not identically zero only for the Bianchi class B:
\begin{eqnarray*}
C^{d}_{bc}=\varepsilon_{bce}\nu^{ed}+a_b \delta^d_c - a_c \delta^d_b.
\end{eqnarray*}
\subsection{3+1 Decomposition of the Einstein equations}

We will use the 3+1 decomposition of the Einstein equations as made in \cite{RA}. Comparing our metric (\ref{mt}) with (2.28) of \cite{RA}
 we have that $\alpha=1$ and $\beta^a=0$ which means that the lapse function is the identity and the shift vector vanishes. There the abstract index 
notation is used. We can interpret the quantities as being frame components. For details we refer to chapter 2.3 of \cite{RA}. There are different
 projections of the energy momentum tensor which are important 
\begin{eqnarray*}
   \rho&=&T^{00},\\
     j_a&=&T_a^0,\\
  S_{ab}&=&T_{ab},
\end{eqnarray*}
where $\rho$ is the energy density and $j_a$ is the matter current.

The second fundamental form $k_{ab}$ can be expressed as:
\begin{eqnarray}\label{a}
 \dot{g}_{ab}=-2 k_{ab}.
\end{eqnarray}
The Einstein equations:
\begin{eqnarray}\label{EE}
 \dot{k}_{ab}=R_{ab}+k~k_{ab}-2k_{ac}k^c_b-8\pi(S_{ab}-\frac{1}{2}g_{ab}S)-4\pi\rho g_{ab},
\end{eqnarray}
where we have used the notations $S =g^{ab}S_{ab}$, $k = g^{ab}k_{ab}$, and $R_{ab}$ is the Ricci tensor of the three-dimensional metric.
The evolution equation for the mixed version of the second fundamental form is (2.35) of \cite{RA}:
\begin{eqnarray}\label{MV}
 {{\dot{k}}^a}_{b}=R^a_b+k~k^a_b - 8 \pi S^a_b + 4 \pi \delta^a_b (S -\rho).
\end{eqnarray}
From the constraint equations since $k$ only depends on the time variable we have that:
\begin{eqnarray}\label{CE1}
 R-k_{ab}k^{ab}+k^2&=&16\pi\rho,\\
\label{CE2}
 \nabla^a k_{ab}& = & 8 \pi j_b,
\end{eqnarray}
where $R$ is the Ricci scalar curvature.
Taking the trace of (\ref{MV}) and using (\ref{CE1}) to eliminate the energy density we obtain:
\begin{eqnarray}\label{in}
 \dot k=\frac{1}{4}(k^2+R+3k_{ab}k^{ab})+4\pi S.
\end{eqnarray}

\subsection{Time origin choice and new variables}\label{tc}

Now with the 3+1 formulation our initial data are $(g_{ij}(t_0), k_{ij}(t_0),f(t_0))$, i.e. a Riemannian metric, a second fundamental form and the distribution function
of the Vlasov equation, respectively, on a three-dimensional manifold $S(t_0)$. This is the initial data set at $t=t_0$ for the Einstein-Vlasov system.
We know that future geodesic completeness holds \cite{GC}. We assume that $k < 0$ for all time following \cite{CCH}. This enables us, since one can always add an arbitrary constant 
to the time origin, to set w.l.o.g. 
\begin{eqnarray}\label{Time}
 t_0=-3/k(t_0).
\end{eqnarray}
The reason for this choice will become clear later and is of technical nature. We will now introduce several new variables in order to use the ones which are common 
in cosmology. We can decompose the second fundamental form introducing $\sigma_{ab}$ as the trace-free part:
\begin{eqnarray*}
 k_{ab}=\sigma_{ab}-H g_{ab}.
\end{eqnarray*}
Using the Hubble parameter:
\begin{eqnarray*}
H=-\frac{1}{3}k,
\end{eqnarray*}
we define:
\begin{eqnarray*}
\Sigma_a^b=\frac{\sigma_a^b}{H},
\end{eqnarray*}
and
\begin{eqnarray}
&&\Sigma_{+}=-\frac12(\Sigma_{2}^2+\Sigma_{3}^3), 
\\
&& \Sigma_{-}=-\frac{1}{2\sqrt{3}}(\Sigma_{2}^2-\Sigma_{3}^3). 
\end{eqnarray}
Define also:
\begin{eqnarray*}
\Omega=8\pi \rho/3H^2,\\
\label{q} q=-1-\frac{\dot{H}}{H^2}.
\end{eqnarray*}
From (\ref{CE1}) we obtain the constraint equation:
\begin{eqnarray*}
 \frac{1}{6H^2}(R-\sigma_{ab}\sigma^{ab})=\Omega-1,
\end{eqnarray*}
and from (\ref{in}) the evolution equation for the Hubble variable:
\begin{eqnarray}\label{H-1}
\partial_t(H^{-1})=\frac{3}{2}+\frac{1}{12}\left(\frac{R}{H^2}+\frac{3}{H^2}\sigma_{ab}\sigma^{ab}\right)+\frac{4\pi S}{3H^2}.
\end{eqnarray}
Combining the last two equations with (\ref{MV}) we obtain the evolution equations for $\Sigma_-$ and $\Sigma_+$:
\begin{eqnarray}
\label{Bla1}&&\dot{\Sigma}_+=H\left[\frac{2R-3(R^2_2+R^3_3)}{6H^2}-\left(3+\frac{\dot{H}}{H^2}\right)\Sigma_++\frac{4\pi}{3H^2}(3S^2_2+3S^3_3-2S)\right],\\
\label{Bla2}&&\dot{\Sigma}_-=H\left[\frac{R_3^3-R^2_2}{2\sqrt{3}H^2}-\left(3+\frac{\dot{H}}{H^2}\right)\Sigma_-+\frac{4\pi(S^2_2-S^3_3)}{\sqrt{3}H^2}\right].
\end{eqnarray}
The equation for the non-diagonal elements
\begin{eqnarray}
\dot{\Sigma}_a^b=H\left[\frac{R^b_a}{H^2}-\left(3+\frac{\dot{H}}{H^2}\right){\Sigma}_a^b-\frac{8\pi S^b_a}{H^2}\right].
\end{eqnarray}
\subsection{Vlasov equation with Bianchi symmetry}
We assume that the distribution function has the same symmetry as the geometry, thus in our left-invariant frame $f$ will not depend on $x^a$. 
Moreover it turns out that the equation simplifies if we express $f$ in terms of $p_i$ instead of $p^{i}$. This can be done using the mass shell relation. Because of our special choice of frame the metric has the simple form (\ref{mt}), thus the first three terms of (\ref{con}) vanish. 
Due to the fact that we are contracting and  the antisymmetry of the structure constant we finally arrive at:
\begin{eqnarray}\label{ve}
 \frac{\partial f}{\partial t}+(p^0)^{-1}C^d_{ba}p^{b}p_{d}\frac{\partial f}{\partial p_a}=0.
\end{eqnarray}
From (\ref{ve}) it is also possible to define the characteristic curve $V^a$, however the
expression simplifies when considering the characteristics $V_a$:
\begin{eqnarray}\label{charak}
 \frac{dV_a}{dt}=(V^0)^{-1}C^d_{ba}V^bV_{d},
\end{eqnarray}
for each $V_i(\bar{t})=\bar{p}_i$ given $\bar{t}$. For the rest of the paper, the capital $V_i$ indicates that $p_i$ is parameterized by the coordinate time $t$.

Note that if we define:
\begin{eqnarray}\label{VVV}
 V=g^{ij}V_iV_j,
\end{eqnarray}
due to the antisymmetry of the structure constants we have with (\ref{charak}):
\begin{eqnarray}\label{ha}
\frac{dV}{dt}=\frac{d}{dt}(g^{ij})V_iV_j.
\end{eqnarray}
Let us also write down the components of the energy momentum tensor in our frame:
\begin{eqnarray}
&&T_{00}=\int f(t,p^{a}) p^0 \sqrt{g}dp^1 dp^2 dp^3,\\
\label{emt2}&&T_{0j}=-\int f(t,p^{a}) p_j \sqrt{g}dp^1 dp^2 dp^3,\\
\label{emt3}&&T_{ij}=\int f(t,p^{a}) p_i p_j (p^0)^{-1}\sqrt{g}dp^1 dp^2 dp^3.
\end{eqnarray}
\section{Isotropic spacetimes with collisionless matter}\label{sap}
Let us start with some considerations concerning isotropic spacetimes. This section is in a sense independent of the following ones, but it may add some intuition
and motivation to the study of Bianchi V spacetimes. The Robertson-Walker metric is as follows:
\begin{eqnarray}
 g_{\alpha\beta}=-dt^{2}+a^{2}(t)\left[\frac{dr^{2}}{1-Kr^{2}}+r^2(d\theta^{2}+\sin^{2}\theta d\phi^{2})\right],
\end{eqnarray}
where $a$ is the (positive) scale factor and $K$ is the curvature parameter of the hypersurfaces of constant cosmological time. 
The cases $K=-1,0,1$ are the hyperboloid (open, special case of Bianchi V and VII$_h$ with $h\neq0$),
flat space (special case of Bianchi I and Bianchi VII$_0$) and the 3-sphere (closed, special case of Bianchi IX)
respectively. The isotropy forces the energy momentum tensor to have the algebraic form $T_{\alpha \beta}=\rho u_\alpha u_\beta + P (g_{\alpha\beta}+u_\alpha u_\beta)$ where $\rho$ is the energy density 
and $P$ the pressure. It was shown in \cite{MM3} that the distribution function can be expressed in terms of $F(\mathcal{P}^2)$ where $\mathcal{P}^2=a^2(p_1^2+p_2^2+p_3^2)$ such that
the energy density and the pressure then take the following form:
\begin{eqnarray*}
&&\rho(t)=\frac{4\pi}{a(t)^4}\int_0^\infty\mathcal{P}^2F(\mathcal{P}) \sqrt{a(t)^2+\mathcal{P}^2}d\mathcal{P},\\
&&P(t)=\frac{4\pi}{3a(t)^4}\int_0^\infty \frac{\mathcal{P}^4F(\mathcal{P})}{ \sqrt{a(t)^2+\mathcal{P}^2}}d\mathcal{P}^2.
\end{eqnarray*}
We can see that $\rho$ and $P$ are non-negative and
\begin{eqnarray}\label{EC}
 \rho \geq 3 P.
\end{eqnarray}
The standard Friedmann equations are
\begin{eqnarray*}
\frac{3({\dot{a}}^{2}+K)}{a^{2}}=8\pi \rho,\\
\frac{2\ddot{a}}{a}+\frac{\dot{a}^{2}+K}{a^{2}}=-8\pi P.
\end{eqnarray*}
 Using the Hubble variable which in this case is $H=\frac{\dot{a}}{a}$, the equations can be put in the following manner:
\begin{eqnarray}
\label{1}&&H^2=\frac{8\pi\rho}{3}-\frac{K}{a^2},\\
\label{2}&&\dot{H}=-4\pi (\rho+P)+\frac{K}{a^2},
\end{eqnarray}
where the first equation was used in the second one. 
Differentiating (\ref{1}) and using (\ref{2})
\begin{eqnarray}\label{3}
 \dot \rho= -3H(\rho+P).
\end{eqnarray}
Using (\ref{1}) in (\ref{2}) we obtain
\begin{eqnarray*}
\dot{H}=-[H^2+\frac{4\pi}{3}(3P+\rho)].
\end{eqnarray*}
Clearly 
\begin{eqnarray}\label{1b}
 \dot{H}\leq -H^2.
\end{eqnarray}
This inequality means that $H$ is always decreasing or can become the constant zero. Since $H(t_0)$ is assumed to be positive there are in principle two possibilities. 
Either $H$ remains always positive or vanishes at some finite time. Consider the first case and $K=1$.
Then $a$ is increasing but $\rho$ is decreasing. Thus at some point, for $\rho_c=\frac{3}{8\pi a_c^2}$, in contradiction to the assumption that it is always positive,
the quantity $H$ vanishes.  In particular with (\ref{2})
\begin{eqnarray*}
 \dot{H}(t_c)=-4\pi P(t_c)- \frac{1}{2a_c^2}.
\end{eqnarray*}
Thus $H$ will decrease forever until the metric becomes singular.
\subsection{Late-time behavior $K\leq 0$}
Let us assume now that $K=0$ or $K=-1$.  Looking at (\ref{1})-(\ref{2}) the possibility that $H$ vanishes is only possible asymptotically since $\rho$ and $-\frac{K}{a^2}$
 are otherwise positive. Thus we have that the Universe expands forever, i.e. $a$ tends to infinity. 
 One can obtain also estimates for $a$.
 Summing the equation (\ref{1}) to (\ref{2}) we obtain
\begin{eqnarray*}
\dot{H}+2H^2=4\pi(\frac13\rho-P)-\frac{K}{a^2}.
\end{eqnarray*}
 Thus for $K\leq 0$ we obtain with (\ref{EC}) and (\ref{1b}):
\begin{eqnarray*}
1  \leq -\frac{\dot{H}}{H^2} \leq 2.
\end{eqnarray*}
Integration with respect to $t$ for $t\ge t_0$ and using (\ref{Time}) implies
\begin{eqnarray*}
 \frac{1}{2t}\leq H \leq \frac1t .
\end{eqnarray*}
Thus both $H$ and $\dot{H}$ are uniformly bounded. Using that $H=\frac{\dot{a}}{a}$ and integrating the last equation again with respect to $t$ for $t\ge t_0$ we also have
\begin{eqnarray*}
 \frac{t^{\frac12}}{t_0^{\frac12}}\leq \frac{a}{a(t_0)}\leq \frac{t}{t_0}.
\end{eqnarray*}
Let us consider now the limits of $\rho$ and $P$ for the asymptotic regime of $a$, namely when $a$ tends to infinity.  Following \cite{Hans} we arrive at
\begin{eqnarray*}
 &&\lim_{t \rightarrow  \infty} a^3 \rho=\alpha_2,\\
 &&\lim_{t  \rightarrow \infty} a^5 P=\frac{\alpha_4}{3},
\end{eqnarray*}
where 
\begin{eqnarray*}
 \alpha_n= 4\pi \int_0^\infty \mathcal{P}^nF(\mathcal{P}) d\mathcal{P}^2.
 \end{eqnarray*}
 Moreover
 \begin{eqnarray*}
 &&\lim_{t \rightarrow  \infty}a^2 (a^3 \rho-\alpha_2)=\frac12 \alpha_4,\\
 &&\lim_{t  \rightarrow \infty} a^2(a^5 P-\frac{\alpha_4}{3})=-\frac16 \alpha_6.
\end{eqnarray*}
 Thus for large $t$
 \begin{eqnarray*}
  \rho=\alpha_2 a^{-3}(1+O(a^{-2})),\\
  P=\frac{\alpha_4}{3}a^{-5}(1+O(a^{-2})).
 \end{eqnarray*}
 Using the estimate for $a$ we see that that asymptotically there is a dustlike behavior since
 \begin{eqnarray*}
  \frac{P}{\rho}=O(t^{-1}).
 \end{eqnarray*}
The dustlike behavior towards the future was already established in e.g. Corollary 3 of \cite{EGS}.

\subsection{Instability of $K=0$}
Define a new time variable via
\begin{eqnarray*}
 \frac{dt}{d\tau}=\frac{1}{H}.
\end{eqnarray*}
We then obtain
\begin{eqnarray*}
 \frac{dH}{d\tau}=-(1+q)H\\
 \end{eqnarray*}
with
\begin{eqnarray*}
 q=-\frac{\ddot{a}a}{\dot{a}^2}=\frac{1}{H^2}\frac{4\pi}{3}(\rho+ 3P),
\end{eqnarray*}
and the Hamiltonian constraint reads
\begin{eqnarray*}
 \Omega=1+\frac{K}{H^2a^2}.
\end{eqnarray*}
For $K\leq 0$ we have $\Omega  \in [0,1]$. The evolution equation of $\Omega$ using (\ref{3}) is:
\begin{eqnarray*}
 \frac{d\Omega}{d\tau}=-(\Omega+8\pi P/H^2)(1-\Omega).
\end{eqnarray*}
Since the variable $\Omega$ is bounded the same follows for $P/H^2$ due to (\ref{EC}). This means then that $\frac{d\Omega}{d\tau}$ is bounded. The derivative of
this expression is bounded if the derivative of $P/H^2$ is bounded. This can be achieved using the Vlasov equation,
integrating by parts and using the fact that $f$ has compact support. Note also that the structure constants
are bounded for the cases in consideration here (see Table 6.1 of \cite{RS} for explicit expressions).
Now all the relevant quantities are bounded. Let $\{\tau_n\}$ be a sequence tending to infinity and let $H_n(\tau)=H(\tau+\tau_n)$, $\Omega_n(\tau)=\Omega(\tau+\tau_n)$, 
$(P/H^2)_n(\tau)=(P/H^2)(\tau+\tau_n)$. Using the bounds already listed, the Arzela-Ascoli theorem \cite{Rudin} can be applied. This implies that, after passing to a subsequence,
 $H_n$, $\Omega_n$, $(P/H^2)_n$ converge uniformly on compact sets to a limit $H_{\infty}$, $\Omega_{\infty}$, $(P/H^2)_{\infty}$ respectively. 
 The first derivative of these variables converges to the corresponding derivative of the limits. Using now the fact that $(P/H^2)_{\infty}$ is zero we have:
\begin{eqnarray}\label{11}
 \frac{d\Omega_{\infty}}{d\tau}=-\Omega_{\infty}(1-\Omega_{\infty}).
\end{eqnarray}
If $K=0$ which is equivalent with $\Omega=1$ we have an equilibrium point of $\Omega$. If $K=-1$ then $0<\Omega<1$ and $\frac{d\Omega}{d\tau}<0$ until 
$\Omega$ is zero and the solution converges to the Milne solution. We cannot apply the Arzela-Ascoli theorem if $K=1$ since for instance $\Omega$ will become
unbounded. However looking at the regime where $a$ tends to zero, by the same methods discussed previously we obtain
\begin{eqnarray*}
  \rho=\alpha_3 a^{-4}(1+O(a^{2})),\\
  P=\frac{\alpha_3}{3} a^{-4}(1+O(a^{2})).
 \end{eqnarray*}
Thus in the regime where $a$ tends to zero
\begin{eqnarray*}
 8\pi P/H^2=\Omega(1+O(a^2)).
\end{eqnarray*}
Thus
\begin{eqnarray}\label{22}
 \frac{d\Omega}{d\tau}=(2\Omega+O(a^2))(\Omega-1).
\end{eqnarray}
We see from (\ref{11}) and (\ref{22}) that $K=0$ which means $\Omega=1$ is an unstable equilibrium point.

\section{Asymptotic behavior of Bianchi V spacetimes}
\subsection{Bianchi V spacetimes}

For Bianchi V spacetimes the trace free part of the Ricci tensor is zero (cf. e.g. 1.69 of \cite{ESW}).
Thus we have 
\begin{eqnarray}
 R^i_j=\frac13 R\delta^i_j.
\end{eqnarray}
This has the consequence that the evolution equations for the $\Sigma$ variables simplify considerably:
\begin{eqnarray}
&&\dot{\Sigma}_+=H\left[-\left(3+\frac{\dot{H}}{H^2}\right)\Sigma_+ +\frac{4\pi}{3H^2}(3S^2_2+3S^3_3-2S)\right],\\
&&\dot{\Sigma}_-=H\left[-\left(3+\frac{\dot{H}}{H^2}\right)\Sigma_-+\frac{4\pi(S^2_2-S^3_3)}{\sqrt{3}H^2}\right].
\end{eqnarray}
The equation for the non-diagonal elements is
\begin{eqnarray}
\dot{\Sigma}_a^b=H\left[-\left(3+\frac{\dot{H}}{H^2}\right){\Sigma}_a^b-\frac{8\pi S^b_a}{H^2}\right].
\end{eqnarray}
Moreover
\begin{eqnarray*}
 R=-6a_1a^1=-6a_1^2g^{11}=-6g^{11},
\end{eqnarray*}
since one can choose w.l.o.g. $a_1=1$. Consider the quantity $A=-\frac{R}{6H^2}=\frac{g^{11}}{H^2}$. The evolution equation of $A$ is
\begin{eqnarray*}
 \dot{A}=2H\left[A\left(2\Sigma_+-1-\frac{\dot{H}}{H^2}\right)+\Sigma^1_2\frac{g^{12}}{H^2}+\Sigma^1_3\frac{g^{13}}{H^2}\right].
\end{eqnarray*}
Define $\tilde{A}=1-A$ which is positive due to the constraint equation, then:
\begin{eqnarray}\label{tilde}
 \dot{\tilde{A}}=-2H\left[(1-\tilde{A})\left(2\Sigma_+-1-\frac{\dot{H}}{H^2}\right)+\Sigma^1_2\frac{g^{12}}{H^2}+\Sigma^1_3\frac{g^{13}}{H^2}\right].
\end{eqnarray}
\subsection{The Milne solution}
When one considers the Einstein-Euler system the solutions are future asymptotic to the Milne solution which in our left-invariant
frame has the following spatial form:
\begin{eqnarray}
g_{ij}=t^2 \delta_{ij}.
\end{eqnarray}
In the variables defined, this solution corresponds to $\Sigma^a_b=0$ for all $a,b$ and $A=1$ or $\tilde{A}=0$.
Our aim is a small data result and we want to show that solutions of the Einstein-Vlasov system with Bianchi V symmetry
are asymptotically stable to the Milne solution. 
\subsection{Bootstrap argument}
The argument which will lead us to our main conclusions is a bootstrap argument, a kind of continuous induction argument. The argument will work as follows (see 10.3 of \cite{RA} for a detailed discussion). One has a solution of the evolution equations and assumes that the norm of
that function depends continuously on the time variable. Assuming that one has small data initially at $t_0$, i.e. the norm of our function
is small, one has to improve the decay rate of the norm such that the assumption that $[t_0,T)$ is the maximal interval with $T<\infty$
would lead to a contradiction. This is a way to obtain global existence for small data. In our case global existence is
already clear but if the argument works we also obtain information about how the solution behaves asymptotically which is our goal. The interval we look at is $[t_0,t_1)$ and we will present the estimates assumed in the following. We will need an additional quantity which we will now define.
 We have a number (different from zero) of particles at possibly different momenta
 and we define $\mathrm{P}$ as the supremum of the absolute value of these momenta at a given time $t$:
\begin{eqnarray*}
 \mathrm{P}(t)=\sup \{ \vert p \vert =(g^{ab}p_a p_b)^\frac{1}{2} \vert f(t,p)\neq 0\}
\end{eqnarray*}
A bound on that quantity can be used for estimates on $S/H^2$ as we show now. Consider an orthonormal frame and denote the components
 of the spatial part of the energy-momentum tensor in this frame by $\widehat{S}_{ab}$. The components can be bounded by
\begin{eqnarray*}
\widehat{S}_{ab} \leq \mathrm{P}^2(t) \rho,
\end{eqnarray*}
so we have that
\begin{eqnarray}\label{PPP}
\frac{\widehat{S}}{\rho} \leq 3\mathrm{P}^2.
\end{eqnarray}
 \subsubsection{Bootstrap assumptions for Bianchi V}

\begin{eqnarray*}
|\Sigma_+| &\leq& \Delta_+ (1+t)^{-\frac{3}{2}},\\
 |\Sigma_-|&\leq& \Delta_-(1+t)^{-\frac{3}{2}},\\
|\Sigma^i_j|&\leq& \Delta_{ij} (1+t)^{-\frac{3}{2}}; i\neq j,\\
|\tilde{A}|&\leq& \Delta,\\
\mathrm{P}&\leq& \Delta_p (1+t)^{-\frac{3}{4}}.
\end{eqnarray*}
The $\Delta$ with different indices are positive and small constants. In the following sections
 $\epsilon$ will denote a strictly positive, but arbitrarily small constant. 
\subsubsection{Mean curvature}
Using the evolution equation for the mean curvature 
\begin{eqnarray}
\partial_t(H^{-1})=1 +\frac12 \tilde{A}+\frac{3}{H^2}\sigma_{ab}\sigma^{ab}+\frac{4\pi S}{3H^2}=1+D,
\end{eqnarray}
which implies since $D$ is positive
\begin{eqnarray}
1\leq \partial_t(H^{-1})\leq 1+\epsilon.
\end{eqnarray}
We can integrate the evolution equation using our time origin choice (\ref{Time}) which implies
\begin{eqnarray}
H=t^{-1}\frac{1}{1+It^{-1}},
\end{eqnarray}
where
\begin{eqnarray}
I=\int^t_{t_0} D(s) ds.
\end{eqnarray}
$D$ is positive  thus we can conclude
\begin{eqnarray}
t^{-1}(1-\epsilon) \leq H \leq t^{-1},
\end{eqnarray}
or
\begin{eqnarray*}
H=t^{-1}(1+O(\epsilon)).
\end{eqnarray*}
\subsubsection{Estimate of the metric and $\mathrm{P}$}
For a matrix $A$ its norm can be defined as:

\begin{eqnarray*}
\|A\|=\sup\{ |A x|/|x|: x\ne 0\}.
\end{eqnarray*}

Let $B$ and $C$ be $n\times n$ symmetric matrices with $C$ positive definite. It is possible to define a relative norm by:

\begin{eqnarray*}
\|B\|_C=\sup\{ |B x|/|C x|: x\ne 0\}.
\end{eqnarray*}

Clearly:

\begin{eqnarray*}
\|B\| \le \|B\|_C \|C\|.
\end{eqnarray*}

It also true that:

 \begin{eqnarray}\label{trick}
\|B\|_C \leq \sqrt{ \tr(C^{-1}BC^{-1}B)}.
 \end{eqnarray}
Using (\ref{trick}) we obtain in the sense of quadratic forms:

\begin{eqnarray}\label{si}
\sigma^{ab}\leq(\sigma_{cd}\sigma^{cd})^{\frac{1}{2}}g^{ab}.
\end{eqnarray}
Define
\begin{eqnarray*}
\bar{g}^{ab}=t^{2}g^{ab}.
\end{eqnarray*}
Then
\begin{eqnarray*}
\frac{d}{dt}(t^{-\gamma}\bar{g}^{ab})= t^{-\gamma-1}\bar{g}^{ab}(-\gamma+2)+2t^{-\gamma+2}(\sigma^{ab}-Hg^{ab}).
\end{eqnarray*}
where we have introduced for technical reasons a small positive parameter $\gamma$. Using now the inequality (\ref{si})

\begin{eqnarray}\label{dec}
\frac{d}{dt}(t^{-\gamma}\bar{g}^{ab})\leq t^{-\gamma-1}\bar{g}^{ab}[-\gamma+2+2tH((H^{-2}\sigma_{cd}\sigma^{cd})^{\frac{1}{2}}-1)].
\end{eqnarray}
Using the equation (\ref{dec}) and the estimate of $H$
\begin{eqnarray*}
\frac{d}{dt}(t^{-\gamma}\bar{g}^{ab})\leq t^{-\gamma-1}\bar{g}^{ab}[-\gamma+2+2(1+O(\epsilon))((H^{-2}\sigma_{cd}\sigma^{cd})^{\frac{1}{2}}-1)].
\end{eqnarray*}
We obtain decay for the metric (in the sense of quadratic forms) provided that $(H^{-2}\sigma_{cd}\sigma^{cd})^{\frac{1}{2}}\leq1$. 
This holds due to our bootstrap assumptions. Thus we have
\begin{eqnarray*}
 g^{ab} \leq t^{-2} t_0^{2}g^{ab}(t_0).
\end{eqnarray*}
This implies that the components of the metric are also bounded by some constant $C(t_0)$ which depends on the terms of $g^{ab}(t_0)$.
Consider now
\begin{eqnarray*}\label{cons}
\dot{g}^{bf}=2H(\Sigma^b_a-\delta^b_a)g^{af}.
\end{eqnarray*} 
Since the metric components are bounded the non-diagonal terms will contribute only with an $\epsilon$. Thus we have for every
component $g^{ij}$ (no summation over the indices in the following equation):
\begin{eqnarray*}
\dot{g}^{ij}=2H(\Sigma^i_i-1+\epsilon)g^{ij}\leq 2H (\max(\Sigma^i_i)-1+\epsilon)g^{ij}= 2H(-1 +\epsilon)g^{ij}.
\end{eqnarray*}
Using now the estimate of $H$
\begin{eqnarray}\label{dmetric}
\dot{g}^{ij}\leq t^{-1}(-2+\epsilon)g^{ij}.
\end{eqnarray}
One can conclude that
\begin{eqnarray*}
 \| g^{-1}\| = O(t^{-2+\epsilon}).
 \end{eqnarray*}
Note that a corresponding estimate for  $\| g \|$ can also be obtained.
From  (\ref{dmetric})
\begin{eqnarray*}
\dot{V}=\dot{g}^{bf}V_bV_f\leq t^{-1}(-2+\epsilon)V,
\end{eqnarray*}
which means that
\begin{eqnarray*}
V=O(t^{-2+\epsilon}),
\end{eqnarray*}
which gives us for $\mathrm{P}$:
\begin{eqnarray*}
\mathrm{P}=O(t^{-1+\epsilon}),
\end{eqnarray*}
which leads us to
\begin{eqnarray*}
\frac{S}{H^2}=O(t^{-2+\epsilon}).
\end{eqnarray*}
\subsubsection{Closing the bootstrap argument}
From these estimates it follows immediately that
\begin{eqnarray*}
&&\Sigma_-=O(t^{-2+\epsilon}),\\
&&\Sigma_+=O(t^{-2+\epsilon}),\\
&&\Sigma_i^j=O(t^{-2+\epsilon}); i \neq j.
\end{eqnarray*}
What remains to be improved to close the bootstrap argument is $\tilde{A}$. From equation (\ref{tilde}) one can compute that the first order terms look as follows
\begin{eqnarray*}
 \dot{\tilde{A}}=-2H[2\Sigma_+ + \frac32 \tilde{A}].
\end{eqnarray*}
Choosing $\vert \Sigma_+ \vert < \frac34 \tilde{A}$ initially it follows that
\begin{eqnarray*}
 \tilde{A}=O(t^{-\epsilon}),
\end{eqnarray*}
which is an improvement. Thus with a bootstrap argument (cf. \cite{E3},\cite{E4} for a more details) we can conclude:
\begin{prop*}
Consider any $C^{\infty}$ solution of the Einstein-Vlasov system with Bianchi V symmetry and with $C^{\infty}$
 initial data. Assume that $|{\Sigma}_+(t_0)|$, $|\Sigma_-(t_0)|$, $|\Sigma^i_j(t_0)|$ for all $i\neq j$, $\tilde{A}(t_0)$,  and $\mathrm{P}(t_0)$ are sufficiently small. Then at
 late times there exists an arbitrarily small strictly positive constant  $\epsilon$ such the following estimates hold:
\begin{eqnarray*}
 H(t)&=&t^{-1}(1+O(t^{-\epsilon})),\\
\Sigma_+&=&O(t^{-2+\epsilon}),\\
\Sigma_-&=&O(t^{-2+\epsilon}),\\
\Sigma^i_j&=&O(t^{-2+\epsilon}); i\neq j, \\
\tilde{A}&=&O(t^{-\epsilon}),\\
\mathrm{P}(t)&=&O(t^{-\frac{1}{2}+\epsilon}).
\end{eqnarray*}
\end{prop*}
\section{Main results}
\subsection{Arzela-Ascoli}
Until now we have obtained estimates which show that the decay rates of the different variables
are up to an $\epsilon$ the decay rates one obtains from the linearization. We want to use the Arzela-Ascoli theorem. 
All of the relevant variables and their derivatives are bounded. 
The variables $\Sigma_-$, $\Sigma_+$, $\Sigma_i^j$ with $i\neq j$ and $A$ are bounded uniformly due
 to our estimates and the constraint equation. The Hubble variable $H$ and its derivative is bounded as well.
We have also obtained with the bootstrap argument that $\mathrm{P}$, which is non-negative, decays which means that $S/H^2$ is bounded. 
From the estimates obtained it is clear that $g^{ab}$ and its derivative are bounded. 
Now having a look at the equations we see that the derivatives of $\Sigma_-$, $\Sigma_+$, $\Sigma_i^j$ with $i\neq j$ and $A$
 are also bounded uniformly.
We can bound the derivative of $S$ with the Vlasov equation and integrating by parts. For a more detailed analysis we refer to \cite{E3}.
This means that also the second derivatives of $\Sigma_-$, $\Sigma_+$, $\Sigma_i^j$ with $i\neq j$ and $A$ and $H$ are bounded.
Now all the relevant quantities are bounded and the Arzela-Ascoli theorem can be applied. As a consequence:
\begin{eqnarray*}
 H_{\infty}=t^{-1},
\end{eqnarray*}
where now the sequence has been taken in the variable $t$. We thus obtain the optimal decay rates for the metric and for its derivative. This implies that we obtain the optimal decay
 rates for $\mathrm{P}$. Since $S/H^2$ is zero asymptotically we obtain the optimal estimates for $\Sigma_-$, $\Sigma_+$, $\Sigma_i^j$ with $i\neq j$. Let us summarize:
\begin{thm*}
Consider any $C^{\infty}$ solution of the Einstein-Vlasov system with Bianchi V symmetry and with $C^{\infty}$
 initial data. Assume that $|{\Sigma}_+(t_0)|$, $|\Sigma_-(t_0)|$, $|\Sigma^i_j(t_0)|$ for all $i\neq j$, $|A(t_0)-1|$,  and $\mathrm{P}(t_0)$ are sufficiently small. Then at
 late times there exists an arbitrarily small strictly positive constant  $\epsilon$ such that  the following estimates hold:
\begin{eqnarray*}
 H(t)&=&t^{-1}(1+O(t^{-\epsilon})),\\
\Sigma_+&=&O(t^{-2}),\\
\Sigma_-&=&O(t^{-2}),\\
\Sigma^i_j&=&O(t^{-2}); i\neq j, \\
\tilde{A}&=&O(t^{-\epsilon}),\\
\mathrm{P}(t)&=&O(t^{-\frac{1}{2}}),
\end{eqnarray*}

\end{thm*}
\subsection{Consequences}
By the same methods as in \cite{EN}
\begin{cor}
Consider the same assumptions as in the previous theorem. Then
\begin{eqnarray}\label{g}
g_{ij}=t^2(G_{ij}+O(t^{-1})),
\end{eqnarray}
with $G_{ij}$ independent of time. The corresponding result for the inverse metric also holds.
\begin{eqnarray*}
 p= \frac13 +O(t^{-2}),
\end{eqnarray*}
where $p$ are the Kasner exponents.
\end{cor}
We see that as in Bianchi I there is isotropization, in this case even faster.
Considering the derivative of $Vt^2$ one can see that $Vt^2$ converges to $C+O(t^{-1})$. This in combination with (\ref{g})
leads us to
\begin{cor}
Consider the same assumptions as in the previous theorem. Then:
\begin{eqnarray*}
|V_i| \leq C+O(t^{-1}).
\end{eqnarray*}
\end{cor}
Since $f(t_0,p)$ has compact support on $p$, we obtain that there exists a constant C such that:
\begin{eqnarray*}
 f(t,p)=0 \ \ \ |p_i| \geq C.
\end{eqnarray*}
Let us denote by $\hat{p}$ the momenta in an orthonormal frame. Since $f(t,\hat{p})$ is constant
along the characteristics we have:
\begin{eqnarray*}
 |f(t,\hat{p})|\leq \Vert f_0 \Vert =\sup \{|f (t_0, \hat p)| \}.
\end{eqnarray*}
Putting these facts together we arrive at the estimates which we summarize in the following:
\begin{cor}
Consider the same assumptions as in the previous theorem. Then
\begin{eqnarray*}
&& \rho= O(t^{-3}),\\
&& j_i = O(t^{-4}),\\
&&S_{ij}=O(t^{-5}).
\end{eqnarray*}
\end{cor}
We see that also here we have a dustlike behavior.
\section{Outlook}
We have shown that the spatially flat Friedmann solution with collisionless matter is unstable within the class of isotropic solutions. The asymptotic behavior
of forever expanding isotropic
solutions is determined. They all tend to the Milne solution. It would be of interest to see whether in the Boltzmann case a similar statement can be made. There has been some recent progress in the spatially
flat case where global existence was shown for a certain class of collision cross sections of the hard potential type  \cite{LeeRen2}.  Global existence and late-time behavior was shown recently in the context of Newtonian cosmology \cite{Lee1} and for the spatially flat case  in \cite{Lee2}.

For not necessarily isotropic solutions of the Einstein-Vlasov system with Bianchi V symmetry we have shown future non-linear stability
and that these solutions tend asymptotically to the Milne solution. Bianchi V spacetimes are the simplest type within class B Bianchi spacetimes.
All models of Bianchi B types IV, V, VI$_h$ and VII$_h$ coupled to a fluid are asymptotically self-similar solutions, in particular, they tend to plane wave states. 
In observational cosmology there has been some recent interest in the Bianchi VII$_h$ type. Although there is no direct evidence for a physical Bianchi VII$_h$
model, the Planck data according to \cite{Planck} provide evidence supporting a phenomenological Bianchi VII$_h$ component. It should be possible to do a similar
stability analysis to the one presented here which may shed some light on this discussion.
The smallness assumption used is a technical assumption due to the bootstrap argument. From the corresponding analysis of the perfect fluid case, it should be possible to remove this assumption using a different technique.
\section*{Acknowledgments} 
The work was iniated while SB was visiting LA at the Max-Planck-Institute for Gravitational Physics with a stipend of the German Academic Exchange Service 
and while EN was still there, funded by the SFB 647, a project of the German Research Foundation. EN has been funded by the G\"{o}ran Gustafsson Foundation for Research
in Natural Sciences and Medicine and is now funded by the Irish Research Council. He thanks Alan Rendall for several discussions, Hans Ringstr\"{o}m for the opportunity to read the manuscript of \cite{Hans} and acknowledges the hospitality of Dalhousie University where part of the work was done.
\bibliographystyle{unsrt}

\begin{thebibliography}{10}

\bibitem{Hans}
Hans Ringstr{\"{o}}m.
\newblock {\em {On the Topology and Future Stability of the Universe}}.
\newblock Oxford University Press, Oxford, 2013.

\bibitem{EN}
E.~Nungesser.
\newblock {Isotropization of non-diagonal Bianchi I spacetimes with
  collisionless matter at late times assuming small data}.
\newblock {\em Class. Quant. Grav.}, 27:235025, 2010.

\bibitem{EW2}
G.~F.~R. Ellis and J.~Wainwright.
\newblock {Friedmann-Lemaitre universes}.
\newblock In J.~Wainwright and G.~F.~R. Ellis, editors, {\em {Dynamical Systems
  in Cosmology}}. Cambridge University Press, Cambridge, 1997.

\bibitem{Muk}
V.~Mukanov.
\newblock {\em {Physical Foundations of Cosmology}}.
\newblock Cambridge University Press,, 2005.

\bibitem{ColeyHervik}
A.~A. Coley and S.~Hervik.
\newblock {A dynamical systems approach to the tilted Bianchi models of
  solvable type}.
\newblock {\em Class. Quant. Grav.}, 22:579--606, 2005.

\bibitem{HVC}
S.~Hervik, R.~J. van~den Hoogen, and A.~A. Coley.
\newblock {Future Asymptotic Behaviour of Tilted Bianchi models of type IV and
  VII$_h$}.
\newblock {\em Class. Quant. Grav.}, 22:607--634, 2005.

\bibitem{CHL}
A.~A. Coley, S.~Hervik, and W.~C. Lim.
\newblock {Fluid observers and tilting cosmology}.
\newblock {\em Class. Quant. Grav.}, 23:3573--3591, 2006.

\bibitem{CHLM}
A.~A. Coley, S.~Hervik, W.~C. Lim, and M.~A.~H. MacCallum.
\newblock {Properties of kinematic singularities}.
\newblock {\em Class. Quant. Grav.}, 26:215008, 2009.

\bibitem{Anderson}
M.~T. Anderson.
\newblock {On Long-Time Evolution in General Relativity and Geometrization of
  3-Manifolds}.
\newblock {\em Commun. Math. Phys.}, 222:533--567, 2001.

\bibitem{AM}
L.~Andersson and V.~Moncrief.
\newblock {Future complete vacuum spacetimes. The Einstein equations and the
  large scale behavior of gravitational fields}.
\newblock In P~T Chrusciel and H~Friedrich, editors, {\em {The Einstein
  equations and the large scale behavior of gravitational fields}}. Birkhauser,
  Basel, 2004.

\bibitem{AM2}
L.~Andersson and V.~Moncrief.
\newblock Einstein spaces as attractors for the {E}instein flow.
\newblock {\em J. Diff. Geom.}, 89,1:1--47, 2011.

\bibitem{CollinsH}
C.~B. Collins and S.~W. Hawking.
\newblock {Why is the universe isotropic?}
\newblock {\em The Astrophysical Journal}, 180:317--334, 1973.

\bibitem{Martin}
M.~Reiris.
\newblock {General $K=-1$ Friedman-Lema{\^{\i}}tre models and the averaging
  problem in cosmology}.
\newblock {\em Class. Quant. Grav.}, 25:085001, 2008.

\bibitem{CPZ}
A.~A. Coley, N.~Pelavas, and R.~Zalaletdinov.
\newblock {Cosmological solutions in Macroscopic Gravity}.
\newblock {\em Phys. Rev. Lett.}, 95:151102, 2005.

\bibitem{St}
J.~M. Stewart.
\newblock {\em {Non-equilibrium relativistic kinetic theory}}, volume~10 of
  {\em {Lecture Notes in Physics}}.
\newblock Springer, Berlin, 1971.

\bibitem{RA}
Alan~D. Rendall.
\newblock {\em {Partial differential equations in general relativity}}.
\newblock Oxford University Press, Oxford, 2008.

\bibitem{JS}
J.~M. Stewart.
\newblock {\em {Advanced General Relativity}}.
\newblock Cambridge University Press, Cambridge, 1991.

\bibitem{RIV}
Alan~D. Rendall.
\newblock {The Einstein-Vlasov system}.
\newblock In P~T Chrusciel and H~Friedrich, editors, {\em {The Einstein
  equations and the large scale behavior of gravitational fields}}. Birkhauser,
  Basel, 2004.

\bibitem{HaEl}
S.~W. Hawking and G.~F.~R Ellis.
\newblock {\em {The large scale structure of space-time}}.
\newblock Cambridge University Press, Cambridge, 1973.

\bibitem{GC}
Alan~D. Rendall.
\newblock {Global properties of locally spatially homogeneous cosmological
  models with matter}.
\newblock {\em Math. Proc. Camb. Phil. Soc.}, 118:511--526, 1995.

\bibitem{CCH}
Alan~D. Rendall.
\newblock {Cosmic censorship for some spatially homogeneous cosmological
  models}.
\newblock {\em Ann. Phys.}, 233:82--96, 1994.

\bibitem{MM3}
R~Maartens and S.~D. Maharaj.
\newblock {Invariant Solutions of Liouville's Equation in Robertson-Walker
  Space-Times}.
\newblock {\em Gen. Rel. Grav.}, 19,12:1223--1234, 1987.

\bibitem{EGS}
J.~Ehlers, P.~Geren, and R.~K. Sachs.
\newblock {Isotropic solutions of the Einstein-Liouville Equations}.
\newblock {\em J. Math. Phys.}, 9,9:1344--1349, 1968.

\bibitem{RS}
M.~P. Ryan and L.~C. Shepley.
\newblock {\em {Homogeneous relativistic cosmologies}}.
\newblock Princeton University Press, Princeton, 1975.

\bibitem{Rudin}
W.~Rudin.
\newblock {\em {Principles of Mathematical Analysis}}.
\newblock McGraw-Hill Book Company, 1964.

\bibitem{ESW}
G.~F.~R. Ellis, S.~T.~C. Siklos, and J.~Wainwright.
\newblock {Geometry of cosmological models}.
\newblock In J.~Wainwright and G.~F.~R. Ellis, editors, {\em {Dynamical Systems
  in Cosmology}}. Cambridge University Press, Cambridge, 1997.

\bibitem{E3}
E.~Nungesser.
\newblock {Future non-linear stability for reflection symmetric solutions of
  the Einstein-Vlasov system of Bianchi types II and VI$_0$}.
\newblock {\em Ann. Henri Poincare}, 14, 4:967--999, 2013.

\bibitem{E4}
E.~Nungesser.
\newblock {Future non-linear stability for solutions of the Einstein-Vlasov
  system of Bianchi types II and VI$_0$}.
\newblock {\em J. Math. Phys.}, 53:102503, 2012.

\bibitem{LeeRen2}
H.~Lee and A.~D. Rendall.
\newblock {The spatially homogeneous relativistic Boltzmann equation with a
  hard potential}.
\newblock {\em arXiv:1301.0106 [gr-qc]}, 2013.

\bibitem{Lee1}
H.~Lee.
\newblock {Global solutions of the Vlasov-Poisson-Boltzmann system in a
  cosmological setting}.
\newblock {\em J. Math. Phys.}, 54:073302, 2013.

\bibitem{Lee2}
H.~Lee.
\newblock {Asymptotic behaviour of the relativistic Boltzmann equation in the
  Robertson-Walker spacetime}.
\newblock {\em arXiv:1307.5688 [math-ph]}, 2013.

\bibitem{Planck}
P.~A.~R. Ade et~al.
\newblock {Planck 2013 results. XXVI. Background geometry and topology of the
  Universe}.
\newblock {\em arXiv:1303.5086 [astro-ph.CO]}, 2013.

\end{thebibliography}

\end{document}